\documentclass[prx,aps,twocolumn,preprintnumbers,amsmath,amssymb,superscriptaddress,showpacs]{revtex4-2}

\usepackage{graphicx}
\usepackage{epstopdf}
\usepackage{amsmath}
\usepackage{multirow}
\usepackage{xcolor}
\usepackage{ulem}
\usepackage{float}
\usepackage{multirow}
\usepackage{array}

\newcommand{\RNum}[1]{\uppercase\expandafter{\romannumeral #1\relax}}
\usepackage[sort&compress]{natbib}

\begin{document}

\title{Evolution of the superconductivity in pressurized La$_{3-x}$Sm$_x$Ni$_2$O$_7$}
\author{Qingyi Zhong}
\altaffiliation{These authors contribute equally to this work.}
\affiliation{School of Science, Sun Yat-Sen University, Shenzhen, Guangdong 518107, China}
\author{Junfeng Chen}
\altaffiliation{These authors contribute equally to this work.}
\affiliation{School of Physics, Sun Yat-Sen University, Guangzhou, Guangdong 510275, China}
\affiliation{Guangdong Provincial Key Laboratory of Magnetoelectric Physics and Devices, Sun Yat-Sen University, Guangzhou, Guangdong 510275, China}
\author{Zhengyang Qiu}
\affiliation{School of Physics, Sun Yat-Sen University, Guangzhou, Guangdong 510275, China}
\affiliation{Guangdong Provincial Key Laboratory of Magnetoelectric Physics and Devices, Sun Yat-Sen University, Guangzhou, Guangdong 510275, China}
\author{Jingyuan Li}
\affiliation{School of Physics, Sun Yat-Sen University, Guangzhou, Guangdong 510275, China}
\affiliation{Guangdong Provincial Key Laboratory of Magnetoelectric Physics and Devices, Sun Yat-Sen University, Guangzhou, Guangdong 510275, China}
\author{Xing Huang}
\affiliation{School of Physics, Sun Yat-Sen University, Guangzhou, Guangdong 510275, China}
\affiliation{Guangdong Provincial Key Laboratory of Magnetoelectric Physics and Devices, Sun Yat-Sen University, Guangzhou, Guangdong 510275, China}
\author{Peiyue Ma}
\author{Mengwu Huo}
\affiliation{School of Physics, Sun Yat-Sen University, Guangzhou, Guangdong 510275, China}
\affiliation{Guangdong Provincial Key Laboratory of Magnetoelectric Physics and Devices, Sun Yat-Sen University, Guangzhou, Guangdong 510275, China}
\author{Hongliang Dong}
\affiliation{Center for High Pressure Science and Technology Advanced Research, Shanghai 201203, China}
\affiliation{Shanghai Key Laboratory of Material Frontiers Research in Extreme Environments, Institute for Shanghai Advanced Research in Physical Sciences, Shanghai 201203, China}
\author{Hualei Sun}
\homepage{sunhlei@mail.sysu.edu.cn}
\affiliation{School of Science, Sun Yat-Sen University, Shenzhen, Guangdong 518107, China}
\affiliation{Guangdong Provincial Key Laboratory of Magnetoelectric Physics and Devices, Sun Yat-Sen University, Guangzhou, Guangdong 510275, China}
\author{Meng Wang}
\homepage{wangmeng5@mail.sysu.edu.cn}
\affiliation{School of Physics, Sun Yat-Sen University, Guangzhou, Guangdong 510275, China}
\affiliation{Guangdong Provincial Key Laboratory of Magnetoelectric Physics and Devices, Sun Yat-Sen University, Guangzhou, Guangdong 510275, China}

\date{\today}

\begin{abstract}
Motivated by the discovery of superconductivity in bilayer La$_3$Ni$_2$O$_7$ at 80 K and the increased superconducting transition temperature, $T_\text{c}$, up to 92 K in single crystals of La$_2$SmNi$_2$O$_7$ under pressure, we systematically study the effect of Sm doping on the superconductivity and structure of La$_{3-x}$Sm$_x$Ni$_2$O$_7$ (0 $\leq$ x $\leq$ 1.5) under pressure. Experimental investigations in polycrystalline samples reveal that Sm doping monotonically decreases the lattice constants $c$ and $a$, thereby enhancing crystal structure distortion and leading to an evolution of the metallic ground state in La$_3$Ni$_2$O$_7$ to an insulating state in La$_{1.5}$Sm$_{1.5}$Ni$_2$O$_7$. The maximum onset $T_\text{c}$ in compounds $x=0.9$ and 1.5 is 89 K, while the pressure that drives the emergence of superconductivity is higher for higher doping levels. The results suggest that the enhancement of $T_\text{c}$ in La$_{3-x}$Sm$_x$Ni$_2$O$_7$ is mainly affected by the compressed $c$ lattice before saturation, and the structure transition is critical for the emergence of superconductivity. Our experimental results provide insight into the influence of elemental substitution on nickelate superconductors, offering a means to increase the transition temperature further.
\end{abstract}


\maketitle
\section{Introduction}

The discovery of superconductivity at 80 K in pressurized La$_3$Ni$_2$O$_7$ \cite{Wang2024nor,Sun2023,Zhang2024h,Hou2023} and ambient thin films \cite{Ko2025,Zhou2025a,Liu2025s,Hao2025,Osada2025} has provided a new platform to elucidate the mechanism of high-$T_\text{c}$ superconductivity. The superconductivity of La$_3$Ni$_2$O$_7$ has been independently confirmed by multiple research groups, with the highest $T_\text{c}$ of the undoped parent compound reaching approximately 83 K \cite{Zhang2024eff,Wang2024p,Li2024ele,Zhou2025i,Shi2025p,Sakakibara2024t,Ueki2025,Li2025i,Wen2024,Liu2025e}. Albeit, superconductivity was also found in trilayer nickelates La$_4$Ni$_3$O$_{10}$~\cite{Li2024sign,Zhu2024s,Sakakibara2024t,Zhang2025sc,Li2025si,Shi2025a,Li2024str} and Pr$_4$Ni$_3$O$_{10}$~\cite{Huang2024s,Chen2025l,Zhang2025b,Pei2024pressure}, and hybrid nickelate La$_5$Ni$_3$O$_{11}$~\cite{Shi2025s,Li2024de}. The bilayer NiO$_6$ octahedra is a building block of the superconducting Rudllesden-Popper (PR) nickelates. Density functional theory calculations suggest that substituting La$^{3+}$ with smaller-radius rare-earth ($R^{3+}$) ions ($R$ = rare-earth element) in La$_3$Ni$_2$O$_7$ could enhance superconductivity by strengthening interlayer antiferromagnetic spin exchange and increasing hopping integrals, thereby potentially raising $T_\text{c}$ \cite{Pan2024,Geisler2024o,Qin2023,Qin2024}. Consequently, element substitution has been regarded as a promising strategy for expanding the family of high-$T_\text{c}$ nickelate superconductors. However, some theoretical studies indicate that $R$ substitution in La$_3$Ni$_2$O$_7$ may also enhance octahedral distortions and in-plane bond disproportionation. These effects could increase the pressure required for the structural transition from the orthorhombic to the tetragonal phase and simultaneously weaken the superconducting pairing strength \cite{Zhang2023t,Geisler2024o}.

Theoretical studies indicate that complete substitution of $R$ elements in $R_3$Ni$_2$O$_7$ can significantly enhance $T_\text{c}$ \cite{Geisler2024o}, yet the synthesis of highly doped La$_{3-x}R_x$Ni$_2$O$_7$ remains challenging. Experimentally, however, only partially substituted compounds have been synthesized for bulk La$_3$Ni$_2$O$_7$ on the La sites~\cite{Zhang2024eff,Wang2024b,Xu2024,Feng2024,Wang2025c,Li2025a} or the Ni sites~\cite{Liu2025em}. The 1/3 of Pr doped compound La$_2$PrNi$_2$O$_7$ exhibits a maximum $T_\text{c}$ of 82.5 K at 16 GPa \cite{Wang2024b}. Although this $T_\text{c}$ shows no significant enhancement compared to La$_3$Ni$_2$O$_7$, the scanning transmission electron microscopy measurements revealed that partial Pr substitution effectively suppresses intergrowth of different RP phases. Similarly, the 30\% neodymium doping compound La$_{2.1}$Nd$_{0.9}$Ni$_2$O$_7$ shows pressure-induced superconductivity with $T_\text{c}$ around 80 K \cite{Feng2024}. Notably, high-pressure resistivity measurements on La$_2$SmNi$_2$O$_7$ revealed a remarkable $T_\text{c}$ of 92 K at 22 GPa. It was suggested that the $T_\text{c}$ is enhanced due to the increased in-plane distortion of $\Delta=(b-a)/(b+a)$ \cite{Li2025a}. These findings suggest that systematic Sm substitution may provide an effective pathway for further increasing $T_\text{c}$. It becomes imperative to investigate whether further Sm doping facilitates an enhancement of the superconducting transition temperature over 92 K.

In this work, we have successfully synthesized La$_{3-x}$Sm$_{x}$Ni$_2$O$_7$ ($x=0$, 0.3, 0.6, 0.9, 1.2, and 1.5) polycrystalline samples via the sol-gel method and investigated their structural, electronic, and magnetic properties under ambient and high pressures. Ambient-pressure X-ray diffraction results show that Sm doping induces significant reductions in lattice parameters $c$ and $a$ compared to La$_3$Ni$_2$O$_7$. La$_{1.5}$Sm$_{1.5}$Ni$_2$O$_7$ has a more in-plane distorted orthorhombic crystal lattice. Under high pressure, La$_{1.5}$Sm$_{1.5}$Ni$_2$O$_7$ undergoes a structural transition from the orthorhombic $Amam$ structure to a tetragonal $I4/mmm$ structure. The high-pressure transport measurements reveal a maximum $T_\text{c}$ of 89 K at 37 GPa, comparable to that of La$_{2.1}$Sm$_{0.9}$Ni$_2$O$_7$. The results suggest that the superconducting transition temperature is enhanced due to the decreased lattice constants, and hence increased electronic and magnetic couplings. However, the $T_\text{c}$ is saturated after 1/3 doping of Sm. The interlayer couplings may play a more significant role than the intralayer couplings.

\begin{figure*}[t]
\includegraphics[width=0.9\textwidth]{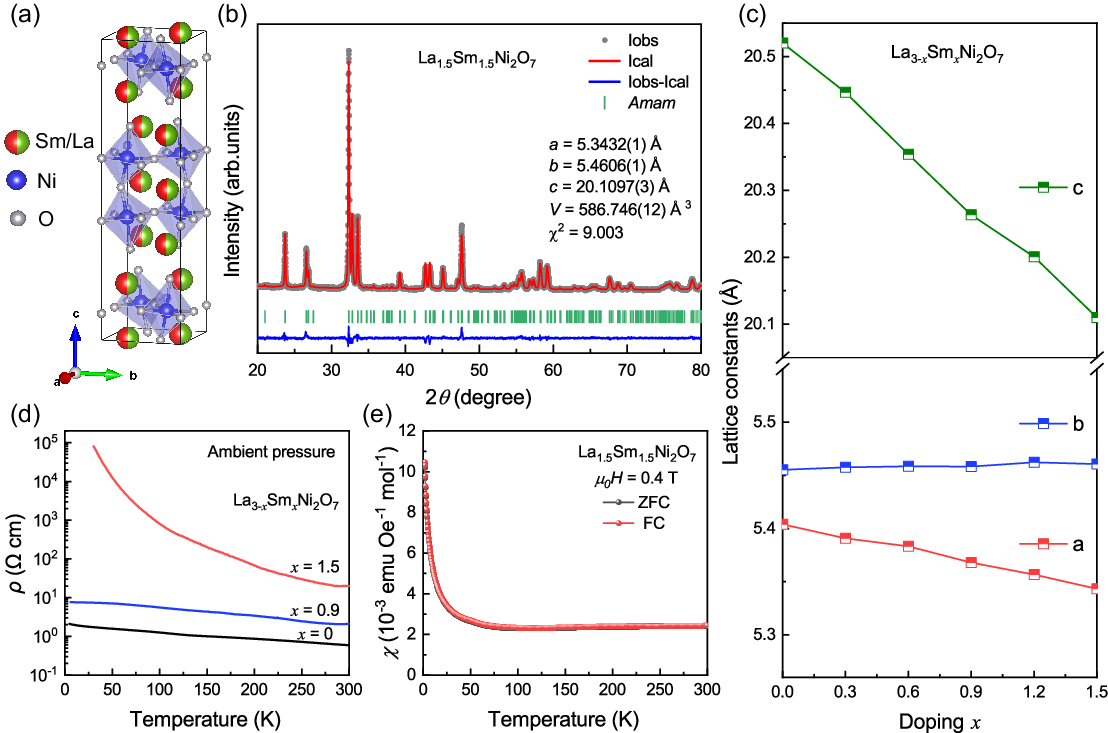}
\caption{(a) Crystal structure of the La$_{1.5}$Sm$_{1.5}$Ni$_2$O$_7$ polycrystalline sample at ambient pressure. (b) Rietveld refinement results of the powder X-ray diffraction pattern, indexed to the orthorhombic $Amam$ space group. (c) The variation of lattice parameters $a$, $b$, $c$ in La$_{3-x}$Sm$_x$Ni$_2$O$_7$ with Sm doping concentration (0 $\leq$ $x$ $\leq$ 1.5). (d) Temperature dependence of electrical resistivity for La$_{3-x}$Sm$_x$Ni$_2$O$_7$ ($x$ = 0, $x$ = 0.9, and $x$ = 1.5) at ambient pressure. (e) Temperature-dependent magnetic susceptibility of La$_{1.5}$Sm$_{1.5}$Ni$_2$O$_7$ measured under 0.4 T magnetic field in zero-field-cooled (ZFC) and field-cooled (FC).}
  \label{fig1}
\end{figure*}

\section{Experimental details}
\subsection{Synthesis of La$_{3-x}$Sm$_{x}$Ni$_2$O$_7$}
Polycrystalline samples of La$_{3-x}$Sm$_{x}$Ni$_2$O$_7$ were synthesized using the sol-gel method following established procedures \cite{Zhang1994}. Stoichiometric amounts of La$_2$O$_3$ (Macklin, 99.99\%), Sm$_2$O$_3$ (Macklin, 99.99\%), and Ni(NO$_3$)$_2$$\cdot$6H$_2$O (Aladdin, 99.99\%) were mixed and dissolved in deionized water, followed by the addition of nitric acid (Aladdin, 70\%) and citric acid (Alfa Aesar, 99+\%). The solution was continuously stirred and heated at 180 $^\circ$C for approximately 4 hours to form a vibrant green gel. The gel was subsequently dehydrated at 200 $^\circ$C for 2 hours, yielding a porous black precursor. To remove organic residues, the precursor was pre-annealed at 800 $^\circ$C for 6 hours. Finally, the obtained powder was thoroughly ground, pressed into pellets, and heated at 1050-1100 $^\circ$C for 24 h to get pure-phase La$_{3-x}$Sm$_{x}$Ni$_2$O$_7$ polycrystals with $x=0$, 0.3, 0.6, 0.9, 1.2, and 1.5.

\subsection{Ambient-pressure characterizations}
The crystal structure and phase purity of polycrystalline samples were characterized by powder X-ray diffraction (XRD) conducted on a diffractometer (Rigabu) with Cu K$\alpha_1$ radiation ($\lambda$ = 1.5406 \AA) at room temperature. Rietveld refinement of the XRD pattern was performed using the Fullprof software package to determine the crystallographic parameters. Electrical transport properties were measured at ambient pressure using a standard four-probe configuration on a Physical Properties Measurement System (PPMS, DynaCool, Quantum Design). Magnetic susceptibility measurements were carried out using a Magnetic Properties Measurement System (MPMS, Quantum Design).

\subsection{High-pressure characterizations}
The polycrystalline La$_{2.1}$Sm$_{0.9}$Ni$_2$O$_7$ and La$_{1.5}$Sm$_{1.5}$Ni$_2$O$_7$ samples were recompressed into square-flakes with dimensions of approximately 55 $\mu$m and a thickness of about 10 $\mu$m separately. High-pressure electrical transport measurements on the samples flake were conducted using a miniature diamond anvil cell (DAC) made of beryllium-copper alloy on a PPMS system. The experiment employed diamond anvils with 300 $\mu$m culets, and a sample chamber with a diameter of approximately 110 $\mu$m was prepared in a cubic boron nitride-epoxy mixed insulating gasket. KBr was used as the pressure-transmitting medium (PTM) to provide a quasi-hydrostatic environment. Electrical measurements were conducted via the van der Pauw method. Pressures below 30 GPa were calibrated by monitoring the fluorescence wavelength shift of a ruby ball in the chamber, while pressures above 30 GPa were calibrated using the Raman spectroscopy of the diamond anvils in the sample region. A discrepancy of approximately 5\%-10\% exists between the two pressure calibration methods.

In situ high-pressure synchrotron powder XRD measurements at 300 K were conducted at Shanghai Synchrotron Radiation Facility using an X-ray wavelength of 0.4834 \AA. A symmetric DAC with 250 $\mu$m culet Boehler-Almax diamonds was utilized, featuring a laser-drilled 160 $\mu$m-diameter sample chamber. Neon gas acted as the pressure-transmitting medium, and chamber pressure was calibrated via XRD of gold under pressure. High-pressure XRD data were first integrated using DIOPTAS (calibrated with CeO$_2$) followed by Rietveld refinement using FullProf software~\cite{Prescher03072015,RODRIGUEZCARVAJAL199355}.

\section{Results}
Figure \ref{fig1} shows investigations of the structural, electronic, and magnetic properties of La$_{3-x}$Sm$_x$Ni$_2$O$_7$ polycrystalline samples at ambient pressure. La$_{1.5}$Sm$_{1.5}$Ni$_2$O$_7$ was measured with emphasis. The crystal structure of La$_{1.5}$Sm$_{1.5}$Ni$_2$O$_7$ is shown in Fig. \ref{fig1}(a), where the 50\% doped rare-earth element Sm randomly occupies La sites. The Rietveld refinement results and corresponding parameters derived from the XRD data at room temperature are presented in Fig. \ref{fig1}(b). La$_{1.5}$Sm$_{1.5}$Ni$_2$O$_7$ exhibits an orthorhombic $Amam$ structure under ambient pressure and remains a pure phase despite the substantial Sm doping. The lattice parameters are $a$ = 5.3432(1) \AA, $b$ = 5.4606(1) \AA, $c$ = 20.1097(3) \AA, and the unit cell volume $V$ = 586.746(12) \AA$^3$. The in-plane Ni-O-Ni bond angle is 155.1(1)$^\circ$, deviating further away 180$^\circ$ compared to La$_3$Ni$_2$O$_7$ \cite{Sun2023,Wang2025c,Li2025a}.  The lattice parameters $a$ and $c$ of La$_{1.5}$Sm$_{1.5}$Ni$_2$O$_7$ decrease significantly with increasing doping, while $b$ shows only slightly increases. This trend holds in the Sm-doped La$_3$Ni$_2$O$_7$ with varying concentrations to $x=1.5$, as shown in Fig. \ref{fig1}(c). The pronounced collapse of the unit cell confirms the successful substitution of half of the La$^{3+}$ ions with smaller radius Sm$^{3+}$ ions.

\begin{figure}[t]
\includegraphics[width=1\columnwidth]{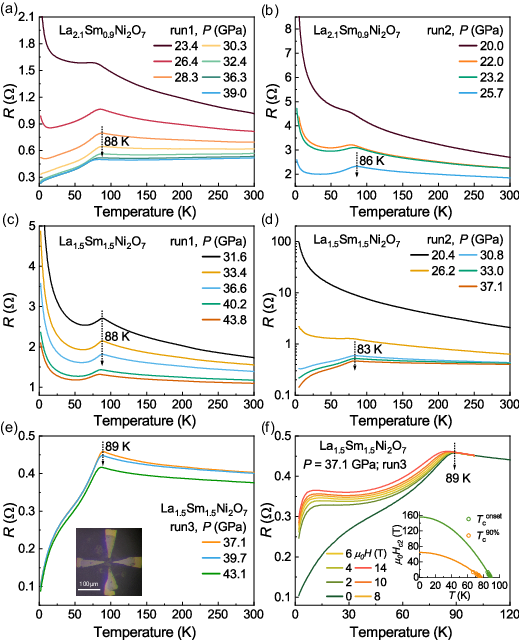}
\caption{(a) The curve of resistance versus temperature for La$_{2.1}$Sm$_{0.9}$Ni$_2$O$_7$ in the pressure range of 23.4 to 39.0 GPa in run 1 and (b) 20.0 to 25.7 GPa in run 2, where the dashed line indicates the superconducting onset transition temperature, $T_c^{\text{onset}}$. (c) Temperature-dependent resistance of La$_{1.5}$Sm$_{1.5}$Ni$_2$O$_7$ in run 1 under pressures ranging from 31.6 to 43.8 GPa, (d) run 2 from 20.4 to 37.1 GPa, and (e) run 3 from 37.1 to 43.1 GPa. The inset shows a photograph of the electrodes for resistance measurements in the high-pressure DAC, with a scale bar of 100 $\mu$m. (f) Magnetic field-dependent resistance in run 3 at 37.1 GPa. The inset presents the upper critical field fitted using $T_c^{\text{onset}}$ and 90\% of $T_c^{\text{onset}}$, where open circles represent experimental data and the solid lines are the results of fitting with the empirical Ginzburg-Landau formula.}
 \label{fig2}
\end{figure}

Figure \ref{fig1}(d) shows the temperature-dependent resistivity of three compositions of La$_{3-x}$Sm$_x$Ni$_2$O$_7$ (where $x$ = 0, $x$ = 0.9, and $x$ = 1.5) under ambient pressure. With increasing doping, the resistivity sequentially rises. Notably,  La$_{1.5}$Sm$_{1.5}$Ni$_2$O$_7$ exhibits insulating behavior, with its resistivity increasing exponentially as temperature decreases and exceeding the instrument detection limit below 30 K. By fitting the resistivity-temperature curve in the 116 - 300 K range, the thermal activation energy for this composition is determined to be 57.4 meV\cite{Liu2023e}. The temperature-dependent magnetic susceptibility of La$_{1.5}$Sm$_{1.5}$Ni$_2$O$_7$ at ambient pressure is shown in Fig. \ref{fig1}(e). The system exhibits purely paramagnetic behavior, as evidenced by the complete overlap of zero-field-cooled (ZFC) and field-cooled (FC) susceptibility curves over the entire temperature range. Fitting the data from 2 - 50 K using the Curie-Weiss law, $\chi(T) = \frac{C}{T - T_{\theta}}$, yields a Curie-Weiss temperature $T_{\theta} = -23.9$ K, Curie constant $C = 0.18$ emu$\cdot$K/mol. This corresponds to an effective magnetic moment $\mu_{eff} = 1.20$ $\mu_B$, significantly larger than the 0.11 $\mu_B$ of undoped La$_3$Ni$_2$O$_7$ but less than the 2.6 $\mu_B$ of La$_{2.1}$Nd$_{0.9}$Ni$_2$O$_7$ \cite{Feng2024}. The fitted effective moment indicates that the introduction of magnetic Sm$^{3+}$ ions introduces localized magnetic moments.

\begin{figure}[t]
\includegraphics[width=1\columnwidth]{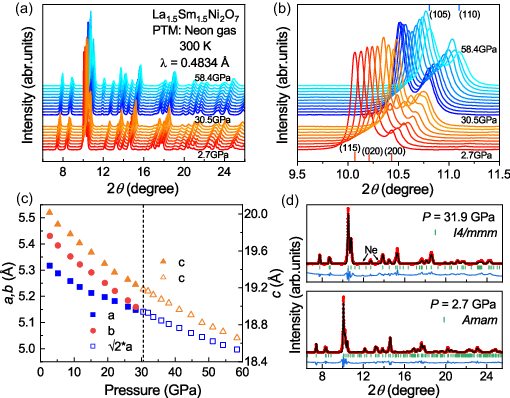}
\caption{(a) High-pressure synchrotron XRD panoramic patterns of the La$_{1.5}$Sm$_{1.5}$Ni$_2$O$_7$ powder sample at room temperature, ranging from 2.7 to 58.4 GPa (wavelength $\lambda$=0.4834 \AA). (b) Detailed evolution of (020)/(200) peaks within the range of 9.5$^\circ$ $\leq$ 2$\theta$ $\leq$ 11.5$^\circ$, showing their pressure-induced merging into a single (110) peak above 30.5 GPa. (c) Variations of lattice parameters $a, b$, and $c$ under pressure. (d) Rietveld refinements at 31.9 GPa and 2.7 GPa using the $I4/mmm$ and $Amam$ space groups, respectively.}
\label{fig3}
\end{figure}

To elucidate the Sm-doping effect on superconductivity, we conducted a comparative measurement of the resistances of polycrystalline La$_{2.1}$Sm$_{0.9}$Ni$_2$O$_7$ and La$_{1.5}$Sm$_{1.5}$Ni$_2$O$_7$ samples under varying pressures. As shown in Fig. \ref{fig2}(a), in run 1, the La$_{2.1}$Sm$_{0.9}$Ni$_2$O$_7$ sample exhibited signs of superconductivity at 23.4 GPa, with an onset transition temperature of 71 K. We note that sharp superconducting transitions of the pressurized RP nickelates only appeared on single crystal samples with a liquid or gas PTM and pollycrystalline samples with a cubic anvil cell which has a larger sample chamber and more homogenious pressure. In Fig. \ref{fig2}, the resistances were measured on polycrystalline samples in a DAC with KBr as the PTM. The drop in resistance closely resembles the behavior of La$_3$Ni$_2$O$_{7-\delta}$ and La$_{2.1}$Nd$_{0.9}$Ni$_2$O$_7$ polycrystalline samples under pressure, revealing a superconducting transition \cite{Zhang2024eff,Feng2024,Ueki2025}. With increasing pressure to 26.4 GPa, the $T_\text{c}$ rapidly rose to 85 K, reaching an optimal value of 88 K at 30.3 GPa before gradually decreasing. In run 2 for another sample with the same composition of $x=0.9$ [Fig. \ref{fig2}(b)], superconductivity emerged at 22.0 GPa.

Figure \ref{fig2}(c) shows the first resistance curve of La$_{1.5}$Sm$_{1.5}$Ni$_2$O$_7$ within the pressure range of $31-36$ GPa. The $T_\text{c}$ remained stable around 88 K, while the upward shift in resistance may stem from grain boundary effects of the powder sample, scattering effect of magnetic Sm$^{3+}$ or oxygen vacancies \cite{Zhang1994}. For run 2 in Fig. \ref{fig2}(d), the $x=1.5$ compound exhibited semiconducting behavior at 20.4 GPa. Upon increasing pressure to 26.2 GPa, resistance dropped sharply by nearly two orders of magnitude, with a transition feature observed at 76 K. Further pressurization to 30.8 GPa led to a steep resistance drop near 83 K, with $T_\text{c}$ eventually stabilizing around 82 K. However, due to the low superconducting volume fraction and pressure inhomogeneity, the transition interval was broad, and zero resistance was not achieved. For run3 in Fig. \ref{fig2}(e), a pronounced resistance decrease was observed in the $37.1-43.1$ GPa range, with the highest $T_\text{c}$ of $\sim$89 K at 37.1 GPa. This transition was progressively suppressed with increasing magnetic field, as shown in Fig. \ref{fig2}(f). According to the empirical Ginzburg-Landau formula $\mu_0H_{c2}(T) = \mu_0H_{c2}(0)\left[1 - \left(T/T_c\right)^2\right]$ fitting, the upper critical magnetic field is 156.1 T at $T_c^{\text{onset}}$ and 64.4 T at 0.9$T_c^{\text{onset}}$. By $\xi_{GL}(0) = \sqrt{\frac{\Phi_0}{2\pi H_{c2}(0)}}$, where $\Phi_0$ is the magnetic flux quantum, equal to 2.068 $\times$ 10$^{-15}$ Wb, their coherence lengths were obtained as 1.45 nm and 2.26 nm, respectively.

To acquire structural information of the La$_{1.5}$Sm$_{1.5}$Ni$_2$O$_7$ polycrystalline under pressure, we conducted synchrotron XRD measurements at room temperature within the pressure range of $2.7-58.4$ GPa. Figure \ref{fig3}(a) presents the synchrotron XRD under varying pressures within the range of 6.0$^\circ$ $\leq$ 2$\theta$ $\leq$ 26.0$^\circ$. Figure \ref{fig3}(b) magnifies the merging of (020) and (200) peaks into a single (110) peak above 30.5 GPa, indicating an orthorhombic ($Amam$)-to-tetragonal ($I4/mmm$) structural transition. The diffraction patterns at low pressures can be indexed using the orthorhombic $Amam$ space group. In contrast, the high-pressure pattern in Fig. \ref{fig3}(d) matches $I4/mmm$ symmetry, confirming the pressure-induced structural transition. Figure \ref{fig3}(c) reveals anisotropic lattice compression, with the lattice parameter $b$ shrinking faster than $a$ in the low-pressure region. The two parameters converge at 30.5 GPa, beyond which a reduces to $1/\sqrt{2}$ of their original values as the crystal symmetry elevates to $I4/mmm$.

\begin{figure}[t]
\includegraphics[width=1\columnwidth]{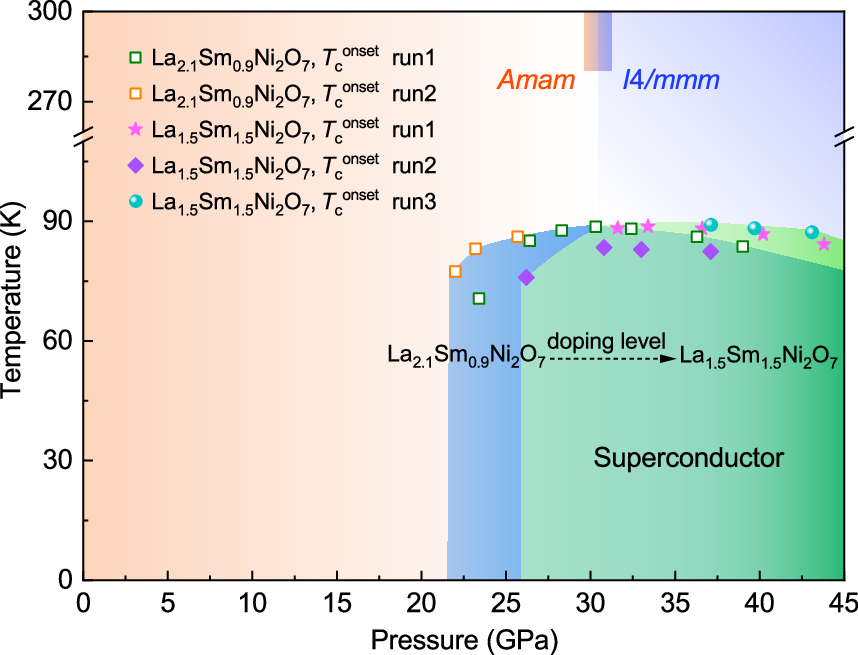}
\caption{The temperature-pressure phase diagram of La$_{2.1}$Sm$_{0.9}$Ni$_2$O$_7$ and La$_{1.5}$Sm$_{1.5}$Ni$_2$O$_7$, where the green and orange squares represent the superconducting onset transition temperature $T_c^{\text{onset}}$ on the resistance-temperature curves for samples La$_{2.1}$Sm$_{0.9}$Ni$_2$O$_7$ run1 and run2, respectively. The pink pentagrams, purple diamonds, and cyan spheres denote the $T_c^{\text{onset}}$ on the resistance-temperature curves for samples La$_{1.5}$Sm$_{1.5}$Ni$_2$O$_7$ run 1, run 2, and run 3, respectively.}
\label{fig4}
\end{figure}

Based on the transport and structural analysis results, we summarized the temperature-pressure phase diagram for La$_{2.1}$Sm$_{0.9}$Ni$_2$O$_7$ and La$_{1.5}$Sm$_{1.5}$Ni$_2$O$_7$ in Fig. \ref{fig4}. The phase diagram include superconducting regions of La$_{2.1}$Sm$_{0.9}$Ni$_2$O$_7$ and La$_{1.5}$Sm$_{1.5}$Ni$_2$O$_7$, and also cover the structural phase transition regions of La$_{1.5}$Sm$_{1.5}$Ni$_2$O$_7$. For the $x=0.9$ and 1.5 polycrystalline samples, superconductivity emerges around 22 GPa and 26 GPa, respectively, with maximum $T_c^{\text{onset}}$ values of 88 K and 89 K, lower than the previous reported single-crystal La$_2$SmNi$_2$O$_7$ \cite{Li2025a}. The onset of superconducting transition temperatures exhibits a broad arch-like feature with pressure, showing a relatively gentle variation. Notably, near 30 GPa at 300 K, the transition from orange to blue marks the structural phase transition of La$_{1.5}$Sm$_{1.5}$Ni$_2$O$_7$ from the orthorhombic $Amam$ phase to the tetragonal $I4/mmm$ phase. This pressure is slightly higher than the 26 GPa at which superconductivity is observed in resistance measurements. Due to the different PTMs in electric transport and structural measurements, pressure uncertainties are expected to exist. A direct correlation between superconductivity and the structural phase transition is possible~\cite{Zhang2024str,Rhodes2024,Wang2025c,Geisler2024s,Wang2024s}. 

As shown in Fig. \ref{fig4}, the $T_c^{\text{onset}}$ values obtained of Sm 50\%-doped La$_{1.5}$Sm$_{1.5}$Ni$_2$O$_7$ did not increase obviously comparing with Sm 30\%-doped La$_{2.1}$Sm$_{0.9}$Ni$_2$O$_7$, but the pressure required for superconductivity to emerge increased. This suggests that for Sm-doped polycrystalline samples, once the doping level reaches a certain threshold and the lattice parameter $c$ is compressed to a specific ratio, the superconducting transition temperature does not exhibit a substantial improvement. In the spin fluctuation pairing mechanism, the $T_c^{\text{onset}}$ is proportional to magnetic exchange couplings, which will also saturate as the lattice is compressed. However, the in-plane distortion of $\Delta$ increases monotonically, distinct from the behavior of $T_c^{\text{onset}}$. It should be noted that $T_c^{\text{onset}}$ for the polycrystalline sample $x=0.9$ is slightly lower than that for the compound of a single crystal $x=1.0$. The decreased $T_c^{\text{onset}}$ may be due to subtle changes in the Sm and O content or the quality of crystallization. The increased pressure for emerging superconductivity with doping should be related to the more distorted structure. Because of the lattice compression, the Ni-O-Ni bond angle deviates further away from 180$^\circ$, which is suggested to be essential for the superconductivity of the RP nickelates.

\section{Conclusion}
We comprehensively investigated high pressure structural evolution and superconductivity of La$_{2.1}$Sm$_{0.9}$Ni$_2$O$_7$ and La$_{1.5}$Sm$_{1.5}$Ni$_2$O$_7$ polycrystalline samples grown by the sol-gel method. Superconductivity emerges in La$_{2.1}$Sm$_{0.9}$Ni$_2$O$_7$ at 22 GPa, while La$_{1.5}$Sm$_{1.5}$Ni$_2$O$_7$ requires a higher critical pressure of 26 GPa. Additionally, La$_{1.5}$Sm$_{1.5}$Ni$_2$O$_7$ undergoes an orthorhombic to tetragonal structural phase transition, which may be related to the emergence of superconductivity. Ambient pressure structural analysis shows that Sm doping exacerbates lattice distortion, yet the maximum $T_c^{\text{onset}}$ values for La$_{2.1}$Sm$_{0.9}$Ni$_2$O$_7$ and La$_{1.5}$Sm$_{1.5}$Ni$_2$O$_7$ show no significant differences, suggesting that excessive doping does not consistently improve the superconducting transition temperature. The enhanced interlayer magnetic exchange couplings may be responsible for the increased $T_c^{\text{onset}}$. Our findings highlight the pivotal role of isovalent doping in modulating the structural characteristics and superconducting properties of nickelates, offering potential insights for improving the transition temperature of high-temperature nickelate superconductors.

\section{Acknowledgments}
Work at SYSU was supported by the National Key Research and Development Program of China (Grants No. 2023YFA1406000, 2023YFA1406500, 2021YFA0718900), the National Natural Science Foundation of China (Grant No. 12425404, 12474137, 12494459010), the Guangdong Basic and Applied Basic Research Funds (Grant No. 2024B1515020040, 2025B1515020008, 2024A1515030030), the Shenzhen Science and Technology Program (Grants No. RCYX20231211090245050), the Guangzhou Basic and Applied Basic Research Funds (Grant No. 2024A04J6417), the CAS Superconducting Research Project (Grant No. SCZX-0101), the Guangdong Provincial Key Laboratory of Magnetoelectric Physics and Devices (Grant No. 2022B1212010008), and Research Center for Magnetoelectric Physics of Guangdong Province (Grant No. 2024B0303390001). We also thank the BL17UM station and the User Experiment Assist System of the Shanghai Synchrotron Radiation Facility for help in characterization.

\bibliography{reference}
\end{document}